\documentclass[twocolumn,showpacs,preprintnumbers,amsmath,amssymb,prb]{revtex4}
\usepackage{graphicx} 
\usepackage{dcolumn} 
\usepackage{bm}
\usepackage{subfigure} 
\usepackage{float} 
\usepackage{color}

\begin{document}

\title{\bf{ Thermodynamics of a two-dimensional frustrated spin-$\frac{1}{2}$ Heisenberg ferromagnet}}

\author{M. H\"{a}rtel}
\author{J. Richter}
\affiliation{Institut f\"{u}r Theoretische Physik, Otto-von-Guericke-Universit\"{a}t Magdeburg, D-39016 Magdeburg, Germany}
\author{D. Ihle}
\affiliation{Institut f\"{u}r Theoretische Physik, Universit\"{a}t Leipzig, D-04109 Leipzig, Germany}
\author{S.-L. Drechsler}
\affiliation{Leibniz-Institut f\"{u}r Festk\"{o}rper- und Werkstoffforschung Dresden, D-01171 Dresden, Germany}

\date{\today}

\begin{abstract}
Using the  spin-rotation-invariant Green's function method we calculate the thermodynamic 
quantities (correlation functions 
$\langle{\bf S}_0{\bf S}_{\bf R}\rangle$, uniform static spin susceptibility $\chi$, 
correlation length $\xi$, and specific heat $C_V$) of the two-dimensional
spin-$1/2$ $J_1$-$J_2$ 
Heisenberg ferromagnet for 
$J_2 < J^c_2 \approx 0.44|J_1|$, where $J^c_2$ is the critical frustrating
antiferromagnetic next-nearest neighbor coupling    
at which the ferromagnetic ground state gives way for a 
ground-state phase with zero
magnetization.
Examining the low-temperature behavior of $\chi$ and $\xi$, in the limit $T \to 0$
both quantities diverge exponentially, i.e., 
 $\chi \propto \exp(b/T)$ 
and $\xi \propto\exp(b/2T)$,
respectively. 
We find a linear decrease in the coefficient $b$ with increasing frustration 
according to 
$ b=-\frac{\pi}{2}\left(J_1+2J_2\right)$, i.e., the exponential divergence
of $\chi$ and $\xi$ 
is present up to $J^c_2$.
Furthermore, we find 
an additional 
low-temperature maximum in the specific heat when approaching 
the critical point, 
$J_2 \to J^c_2$.
\end{abstract}

\maketitle

\section{ Introduction} \label{intro}
Low-dimensional quantum magnets have attracted much attention during the last
decades.\cite{diep04,002} 
They are predestined to study the influence of strong thermal and 
quantum fluctuations. 
Much attention has been paid to the theoretical
investigation of the  
two-dimensional (2D) spin-$1/2$ $J_1$-$J_2$ quantum Heisenberg antiferromagnet 
which may serve as a canonical model to study the interplay  of frustration effects
and quantum fluctuations (see, e.g., Ref.~\onlinecite{j1j2}). 
An additional motivation to study this model comes from the experimental
side. \cite{melzi} 
Very recently, several quasi-2D magnetic 
materials with a ferromagnetic nearest-neighbor (NN) coupling $J_1<0$ 
and a frustrating  
antiferromagnetic next-nearest neighbor (NNN) coupling $J_2>0$ have
been investigated experimentally, 
e.g., Pb$_2$VO(PO$_4$)$_2$, \cite{kaul04,jmmm07,carretta2009,enderle}
(CuCl)LaNb$_2$O$_7$,\cite{kageyama05}
SrZnVO(PO$_4$)$_2$,\cite{rosner09,rosner09a,enderle}
 and
BaCdVO(PO$_4$)$_2$.\cite{nath2008,carretta2009,rosner09} The quite large
frustrating $J_2$ drives these materials out of the ferromagnetic phase.
The experimental findings have
stimulated several theoretical studies of the ground-state and thermodynamic
properties of the $J_1$-$J_2$ model with $J_1<0$ and frustrating
$J_2>0$.
\cite{shannon04,shannon06,sindz07,schmidt07,schmidt07_2,sousa,shannon09,momoi,schulen} 
It was found that the ferromagnetic ground
state
for the spin-$1/2$ model  
breaks down at 
$J_2 = J^c_2 \approx 0.4|J_1|$.
\cite{shannon04,shannon06,sindz07,schmidt07,schmidt07_2,sousa,schulen,dmitriev96}
Note that for the classical model (spin $s \to \infty$) the corresponding transition
point is at $J_2=0.5|J_1|$.
  
On the other hand, some materials considered 
as 2D spin-1/2 ferromagnets, such as 
K$_2$CuF$_4$, Cs$_2$CuF$_4$, Cs$_2$AgF$_4$, La$_2$BaCuO$_5$, 
and Rb$_2$CrCl$_4$,\cite{feldkamp95,feldkamp98,manaka03,tennant05,Kasinathan06} might have also a weak 
frustrating NNN interaction $J_2 < J^c_2$.
In contrast to the previous
investigations of the 2D $J_1$-$J_2$ model with ferrromagnetic
$J_1$ \cite{shannon04,shannon06,sindz07,schmidt07,schmidt07_2,sousa,schulen}  
which have considered predominantly the case of strong frustration, in the
present paper we will focus 
on the region of weak frustration $J_2< J^c_2$.
 
Although for $J_2 <
J^c_2$  the ground state remains ferromagnetic,
the frustrating
$J_2$ may influence the thermodynamics  substantially, in particular near
a zero-temperature transition. 
This has been demonstrated for the one-dimensional (1D) frustrated ferromagnet in
Ref.~\onlinecite{haertel08}, where 
a change in the low-temperature
behavior of the susceptibility and the correlation length 
as well as an additional 
low-temperature maximum
in the specific heat have been found when approaching the zero-temperature critical point
from the ferromagnetic side.

The Hamiltonian of the system considered in this paper is given by
\begin{equation}\label{hamiltonian}
  H=J_1\sum_{\langle i,j\rangle} {\bf S}_i{\bf S}_j+J_2\sum_{[i,j]} {\bf S}_i{\bf S}_j
\; ; \; J_1 < 0 \; ; \; J_2 \ge 0 \; ,
\end{equation}
where $({\bf S}_i)^2=3/4$,  and $\langle i,j\rangle$ denotes NN and $[i,j]$
denotes NNN bonds. 
The unfrustrated 2D ferromagnet ($J_2=0$) has been widely investigated,
e.g., by the modified spin-wave theory,\cite{takahashi} renormalization group
approaches,\cite{kopietz,karchev} the quantum Monte
Carlo method,\cite{ihle_new,manousakis88,chen91,timm2000} 
and by a spin-rotation-invariant second-order Green's function
method (RGM).\cite{kondo,magfeld,antsyg,ihle_new,SSI94}
However, in presence of
frustration ($J_2 > 0$) 
the choice of appropriate methods for studying temperature dependent
quantities of this system is restricted due to frustration and dimensionality. 
For instance, the powerful quantum Monte Carlo method is not applicable due to the minus
sign problem, whereas finite-temperature density matrix renormalization
group studies  are restricted to 1D systems.

So far,  only a few 
theoretical papers deal with the thermodynamics 
of the model for $J_2>0$, see, e.g., Refs.~\onlinecite{shannon04,schmidt07}
and \onlinecite{schmidt07_2},
where finite lattices of $N=16$ and $N=20$ sites are considered. 
As we will see below, for these lattice sizes the finite-size effects at
low temperatures are large. Therefore, methods studying infinite
systems are highly desirable.
Among others, the Green's function method is a powerful tool
to study magnetic systems at finite temperatures, see e.g.
Refs.~\onlinecite{elk,izumov02,froebrich06} and references therein.
A specific variant appropriate for frustated systems is the RGM.   
In the present paper, we use the RGM to investigate thermodynamic properties of the frustrated model
(\ref{hamiltonian}), focusing on 
$J_2 \leq J^c_2 \approx 0.4|J_1|$.
The RGM
has been applied successfully 
to low-dimensional (frustrated) infinite 
quantum spin systems.\cite{haertel08,kondo,rgm_new,j1j2_a,ihlefinite,magfeld,antsyg,ihle_new,SSI94}
In particular, for the 1D spin-$1/2$ Heisenberg ferromagnet 
it was shown that the RGM reproduces 
Bethe-ansatz results.\cite{haertel08,SSI94} Since the RGM can be
formulated also for finite systems, we use full exact diagonalization (ED) 
of finite square lattices of $N=16$ and $N=20$ sites to compare the RGM
with ED data for finite lattices.

The paper is organized as follows: In Sec.~\ref{RGM}
the RGM applied to model (\ref{hamiltonian}) is presented. 
The results for the thermodynamic quantities are discussed in Sec.~\ref{results}. 
Finally, a summary is given in Sec. \ref{summary}.

\section{Rotation-invariant Green's function method}
\label{RGM}
To calculate the thermodynamic quantities of the model (\ref{hamiltonian}), 
we use the RGM \cite{haertel08,kondo,SSI94,rgm_new,j1j2_a} for determining  the transverse dynamic spin susceptibility 
$\chi_{\bf q}^{+-}\left(\omega\right)
=-\langle\langle S_{\bf q}^+;S_{-{\bf q}}^-\rangle\rangle_\omega$, 
where $\langle\langle \ldots;\ldots\rangle\rangle_\omega$ denotes the 
two-time commutator Green's function.\cite{elk} 
Taking the equations of motion up to the second step and supposing spin rotational symmetry,
i.e., $\langle S_i^z\rangle=0$, we obtain 
\begin{equation}\label{a}
\omega^2\langle\langle S_{\bf q}^+;S_{-{\bf q}}^-\rangle\rangle_\omega
=M_{\bf q}+\langle\langle -\ddot{S}_{\bf q}^+;S_{-{\bf
q}}^-\rangle\rangle_\omega \;\; ,
\end{equation}
\begin{equation}
M_{\bf q}=\langle\left[\left[S_{\bf q}^+,H\right],S_{-{\bf
q}}\right]\rangle \;\; , \;\; 
-\ddot{S}_{\bf q}^+=\left[\left[S_{\bf q}^+,H\right],H\right].
\end{equation}
The moment $M_{\bf q}$ is given by the exact expression
\begin{equation}\label{momentMq}
  M_{\bf q}=-8\sum_{k=1,2}J_kC_{1,k-1}\left(1-\gamma_{\bf q}^{(k)}\right) ,
\end{equation}
where $C_{n,m}\equiv C_{\bf R}=\langle S_0^+S_{\bf R}^-\rangle=2\langle S_0^zS_{\bf R}^z\rangle$, 
${\bf R}=n{\bf e}_x+m{\bf e}_y$, $\gamma_{\bf q}^{(1)}=\left(\cos q_x+\cos
q_y\right)/2$, and $\gamma_{\bf q}^{(2)}=\cos q_x\cos q_y$. The second derivative $-\ddot{S}_{\bf q}^+$ is approximated in the 
spirit of the schemes employed in 
Refs.~\onlinecite{haertel08,kondo,SSI94,rgm_new,j1j2_a,ihlefinite,magfeld,antsyg,ihle_new}. 
That is, in $-\ddot{S}_{i}^+$ we adopt the decoupling 
\begin{equation}
S_i^+S_j^+S_k^-
=\alpha_{i,k}\langle S_i^+S_k^-\rangle S_j^++\alpha_{j,k}\langle S_j^+S_k^-\rangle
S_i^+.
\end{equation} 
Following the arguments of Ref.~\onlinecite{haertel08} 
we put $\alpha_{i,k} = \alpha$ in the whole temperature region. We obtain $-\ddot{S}_{\bf q}^+=\omega_{\bf q}^2S_{\bf q}^+$ and
\begin{equation}\label{Greenfunktion}
  \chi_{\bf q}^{+-}\left(\omega\right)=-\langle\langle S_{\bf q}^+;S_{-{\bf q}}^-\rangle\rangle_\omega
=\frac{M_{\bf q}}{\omega_{\bf q}^2-\omega^2},
\end{equation}
with
\begin{align}\label{omegaq}
  \omega_{\bf q}^2&= 2\sum_{k,l(=1,2)}J_kJ_l\left(1-\gamma_{\bf q}^{(k)}\right)\times\nonumber\\
  &\left[K_{k,l}+8\alpha C_{1,k-1}\left(1-\gamma_{\bf q}^{(l)}\right)\right] ,
\end{align}
where $K_{1,1}=1+2\alpha\left(2C_{1,1}+C_{2,0}-5C_{1,0}\right)$, 
$K_{2,2}=1+2\alpha\left(2C_{2,0}+C_{2,2}-5C_{1,1}\right)$, 
$K_{1,2}=4\alpha\left(C_{1,2}-C_{1,0}\right)$, and $K_{2,1}=4\alpha\left(C_{1,0}+C_{1,2}-2C_{1,1}\right)$. 
From the Green's function (\ref{Greenfunktion}) 
the correlation functions $C_{\bf R}=\frac{1}{N}\sum_{\bf q}C_{\bf q}e^{i{\bf q}{\bf R}}$ of arbitrary range ${\bf R}$ are determined by the spectral theorem,
\cite{elk}
\begin{equation}
  C_{\bf q}=\langle S_{\bf q}^+S_{\bf q}^-\rangle=\frac{M_{\bf q}}{2\omega_{\bf q}}\left[1+2n\left(\omega_{\bf
q}\right)\right],
\end{equation}
where 
 $n(\omega_{\bf q})=\left(e^{\omega_{\bf q}/T}-1\right)^{-1}$ is the Bose function.
Taking the
on-site correlator $C_{\bf R=0}$ and using the operator identity 
$S_i^+S_i^-=\frac{1}{2}+S_i^z$ we get the sum rule $C_0=\frac{1}{N}\sum_{\bf q}C_{\bf q}=\frac{1}{2}$. 
The uniform static spin susceptibility $\chi=\lim_{{\bf q}\to 0}\chi_{\bf q}$,
where $\chi_{\bf q}=\chi_{\bf q}(\omega=0)$ and  
$\chi_{\bf q}\left(\omega\right)=\frac{1}{2}\chi_{\bf q}^{+-}\left(\omega\right)$, is given by
\begin{equation}\label{defchi}
\chi=-\frac{2}{\Delta}\sum_{k=1,2}kJ_kC_{1,k-1}, \quad \Delta=\sum_{k,l=1,2}kJ_kJ_lK_{k,l}.
\end{equation}
The correlation length may be calculated by expanding $\chi_{\bf q}$ around 
${\bf q}=0$,\cite{haertel08,kondo,ihle_new} $\chi_{\bf q}=\chi/\left[1+\xi^2\left(q_x^2+q_y^2\right)\right]$. We find
\begin{equation}\label{GlxiQuadrat}
  \xi^2=\frac{2\alpha(J_1^2C_{1,0}+2J_1J_2\left(C_{1,0}+C_{1,1}\right)+4J_2^2C_{1,1})}{\Delta}.
\end{equation}
The ferromagnetic long-range order, occurring at $T=0$ only, is reflected by a 
non-vanishing quantity $C$ (see, e.g.,
Ref.~\onlinecite{kondo})
according to $C_{\bf R}(0)=\frac{1}{N}\sum_{{\bf q}\neq 0}C_{\bf q}(0)e^{i{\bf q}{\bf R}}+C$, 
where $C$, typically called condensation term,\cite{kondo} is connected with the magnetization $m$ by $m^2=3C/2$.
Due to the exact result $C_{\bf R \ne 0}(0)=\frac{1}{6}$, $C_{\bf q}(0)=\frac{M_{\bf
q}(0)}{2\omega_{\bf q}(0)}$ must be 
independent of ${\bf q}$. Examining Eqs.~(\ref{momentMq}) and (\ref{omegaq}), 
this requires $K_{k,l}(0)=0$ which leads to $\alpha(0)=\frac{3}{2}$ 
{and $C_{\bf q}(0)
=\frac{1}{3}$.  
Hence, in case of a ferromagnetic ground state we find at zero temperature
 $\omega_{\bf q}=2\rho_s{\bf q}^2$ ($|{\bf q}|<<1$), $\rho_s
=-\frac{1}{4}(J_1+2J_2)$ (where $\rho_s$ is the
spin-stiffness) and $C_{\bf R}(0)=\frac{1}{3}\delta_{\bf R,0}+\frac{1}{6}$. 
Note that the sum rule $C_0=\frac{1}{2}$ is fulfilled. 
In Eqs.~(\ref{defchi}) and (\ref{GlxiQuadrat}), 
we have $\Delta(0)=0$ so that $\chi$ and $\xi$ diverge as $T\to 0$ 
indicating the ferromagnetic phase transition.

To estimate the zero-temperature transition point $J^c_2$ (see
Sec.~\ref{qpt}), we 
determine the ground-state correlation functions and the ground-state energy 
per site, $E = 3 (J_1 C_{1,0}  + J_2 C_{1,1} )$, for $J_2 > J^c_2$. 
Thereby, we take into consideration the possible existence of a quantum 
collinear phase described by the magnetic ordering vector ${\bf Q} = (\pi, 0)$ or
$(0, \pi)$ and proceed along the lines indicated in
Ref.~\onlinecite{j1j2_a}.

Since we will compare the RGM data with ED data for finite lattices,
see Sec.~\ref{results},  we have to adopt the RGM to finite
$N$.  In this case the quantity $C$ is not related to the 
magnetization and stays nonzero in the whole temperature region. 
In Ref.~\onlinecite{ihlefinite} it was shown that $C=2T\chi/N$.

To evaluate the thermodynamic quantities, a coupled system of six 
(for finite systems: seven) nonlinear algebraic self-consistency equations,
including the sum rule $C_0=\frac{1}{2}$,   
has to be solved numerically to determine the correlators 
$C_{1,0}$, $C_{1,1}$, $C_{2,0}$, $C_{2,1}$, $C_{2,2}, (C)$ and the vertex parameter $\alpha$. 
To solve this system we use Broyden's method,\cite{NR3} which yields the solutions with a 
relative error of about $10^{-8}$ on the average. The momentum integrals are done by Gaussian integration.
To find the numerical solution of the RGM equations for $T> 0$,  we start
at high temperatures and decrease $T$ in small steps. Below a certain
(low) temperature $T_0(J_2)$ no solutions of the RGM equations (except at
$T=0$) could be found,  since the
quantity $\Delta(T,J_2)$ in Eqs.~(\ref{defchi}) and (\ref{GlxiQuadrat}) becomes exponentially small
which leads to numerical instabilities. 

Presenting our results in the next section we put $J_1=-1$.

\section{Results}\label{results}

\subsection{Phase transition at $T=0$} \label{qpt}

In a first step we use the RGM as described above to determine the transition point
$J^c_2$, where the
ferromagnetic ground state gives way for a ground-state phase with zero magnetization. 
\begin{figure}[ht]
  \begin{center}
    \includegraphics[height=6cm]{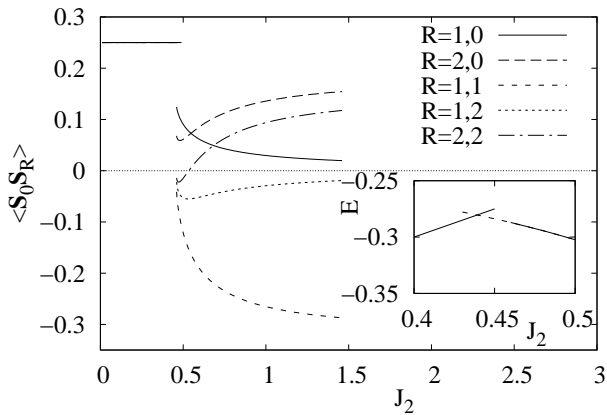}
  \caption{Ground-state spin-spin correlation functions 
$\langle {\bf S}_0 {\bf S}_{\bf R}\rangle$ in dependence on 
frustration $J_2$. (Note that for $J_2 < J^c_2 \approx 0.44$ all $\langle {\bf S}_0 {\bf
S}_{\bf R}\rangle$ coincide.)
Inset: Ground-state energy $E$
versus $J_2$ (solid line: RGM data, dashed line: extrapolated RGM data
to fix the transition point, see also text).
}\label{GSeigenschaften}
  \end{center}
\end{figure}
For that we solve the RGM equations at $T=0$ (i) starting from $J_2=0$ with
increasing $J_2$ and (ii) starting from large $J_2 \gg 1$ with decreasing
$J_2$. In case (i) the RGM equations at $T=0$ can be solved until the classical
transition at $J_2=0.5$.  
In case (ii) we find solutions of the RGM ground-state equations
down to $J_2=0.46$. This value lies certainly above the transition point $J^c_2$.   
To fix $J^c_2$ we may extrapolate the ground-state
energy obtained for case (ii). The crossing point of both energies at
$J_2=0.44$ can
be considered as the RGM estimate of $J^c_2$, see the inset of
Fig.~\ref{GSeigenschaften}. This value is in reasonable agreement with values for $J^c_2$
reported in other papers. \cite{dmitriev96,shannon06,sousa,shannon09,schulen}

The behavior of the spin-spin correlation functions at $T=0$ near the transition point $J^c_2$
is illustrated in Fig.~\ref{GSeigenschaften}.   
For $J_2\gtrsim J^c_2$ there is a noticeable variation in the correlation functions.  
With increasing $J_2$, at $J_2 \gtrsim 1$, the spin-spin correlation
functions 
approach the corresponding values of the limiting case $J_2 \gg 1$. 
This behavior is in qualitative agreement
with data from Lanczos ED and coupled-cluster method.\cite{schulen}
At the critical point the correlation functions jump to the exact results of the ferromagnetic phase, 
i.e., $\langle {\bf S}_0 {\bf S}_{\bf R}\rangle=0.25$. 
 Together with the kink in the energy, this indicates a first-order transition. 
These results confirm the findings in previous
papers.\cite{dmitriev96,shannon06,shannon09} 

\subsection{Thermodynamic properties} 

Now we investigate 
the thermodynamic properties of the model for $J_2 < J^c_2 \approx 0.44$,
where the ground state is ferromagnetic.
\begin{figure}[ht]
  \begin{center}
    \includegraphics[height=6cm]{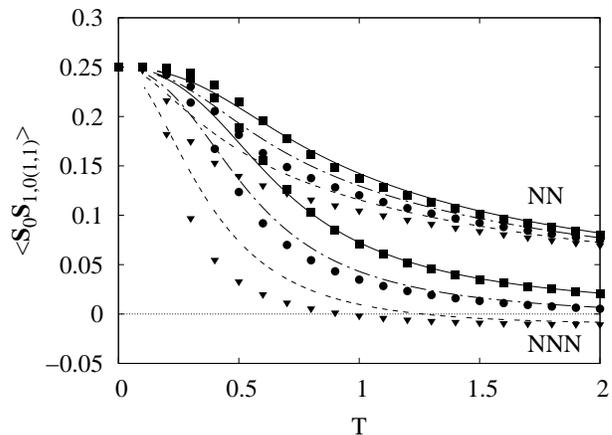}
  \end{center}
  \caption{NN and NNN spin-spin correlation functions 
for $J_2=0$, $0.15$, and $0.3$, from top to bottom, calculated by RGM
for $N\to\infty$ (lines) and ED for $N=20$ 
(filled symbols).}\label{korrfktnen}
\end{figure}
First we consider the temperature dependence of the spin-spin correlation
functions.  The NN
and NNN correlators are shown in Fig. \ref{korrfktnen} for various values of
$J_2$.
For comparison we show also ED data for $N=20$ sites.
The RGM results agree qualitatively with the ED data. 
 With increasing frustration, the correlation functions decrease more
rapidly with temperature. Interestingly, for large 
frustration the NNN correlation function changes its sign at a certain
temperature, e.g., for $J_2=0.3$, at $T\approx 1.25$ (RGM data).
The faster decay of the spin-spin correlators due to frustration
is related
to the decrease in the correlation length $\xi$ with increasing
$J_2$,  see the discussion below.

\begin{figure}[ht]
  \begin{center}
    \includegraphics[height=6cm]{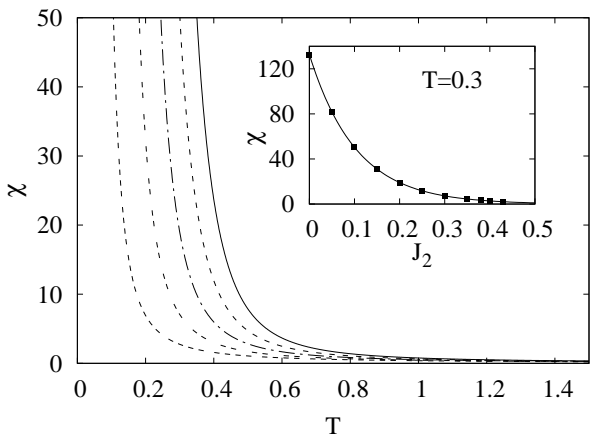}
  \end{center}
  \caption{Uniform static spin susceptibility $\chi$ for
  $J_2=0,0.1,0.2,0.3,0.4$, from right to left. 
The inset shows the susceptibility for $T=0.3$ in dependence on the
frustration $J_2$. The solid line represents an exponential fit, $\chi(T=0.3)
\approx 132.8 \exp(-9.7J_2)$, of the data points (filled squares). 
}\label{chi_T}
\end{figure}
\begin{figure}[ht]
  \begin{center}
    \includegraphics[height=6cm]{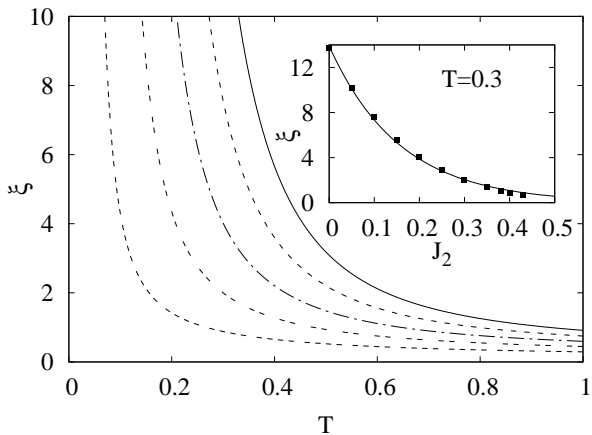}
  \end{center}
  \caption{Correlation length $\xi$ for $J_2=0,0.1,0.2,0.3,0.4$, from right to left. 
The inset shows the correlation length for $T=0.3$ in dependence on the
frustration $J_2$. The solid line represents an exponential fit, $\xi(T=0.3)
\approx 13.9 \exp(-6.4J_2)$, of the data points (filled squares). 
}\label{xi_T}
\end{figure}

The uniform static spin susceptibility $\chi$ and the correlation length
$\xi$ depicted in 
Figs.~\ref{chi_T} and \ref{xi_T} show a similar behavior. Due to the ferromagnetic ground
state, both quantities diverge at $T=0$ exponentially, see below.
With increasing frustration $J_2$ the rapid increase in both quantities is
shifted to lower temperatures. As shown in the insets of
Figs.~\ref{chi_T} and \ref{xi_T}, at a certain fixed temperature both $\chi$ and
$\xi$ decrease rapidly  with $J_2$.

\begin{figure}[ht]
  \begin{center}
    \includegraphics[height=12cm]{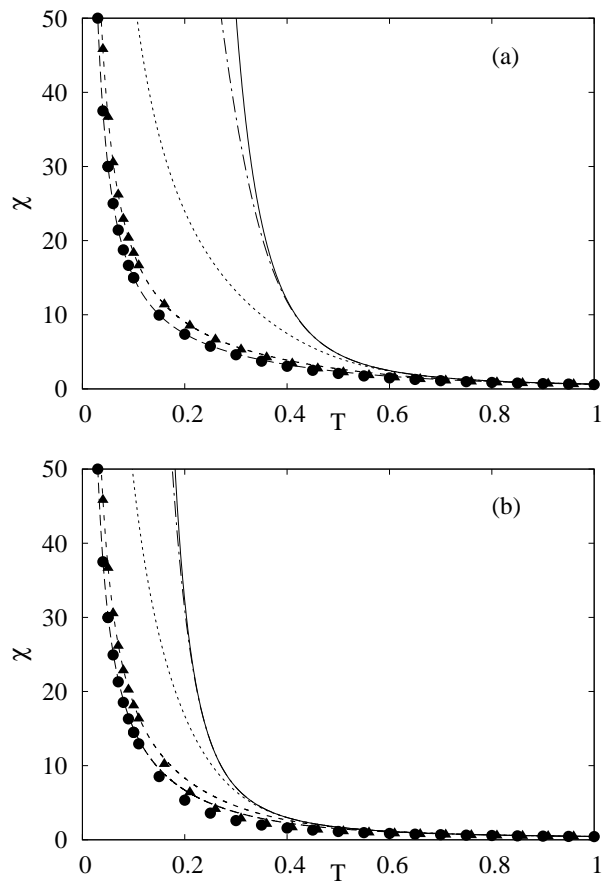}
  \end{center}
  \caption{Uniform static spin susceptibility $\chi$ for 
(a) $J_2=0.1$ and (b) $J_2=0.3$ for $N=16,20,64,400,$ 
thermodynamic limit (lines, RGM) and $N=16,20$ (filled symbols, ED), from left to
right. Note that for $J_2=0.3$ the curves for $N=400$ and $N \to \infty$
almost coincide.  }\label{chiN}
\end{figure}

Since the susceptibility (together with the correlation length) is an
important quantity to analyze the critical properties for $T \to 0$,  we
first
test the quality of our RGM approach in more detail by comparing the RGM data for
$N=16$ and
$N=20$ with ED data for the same system sizes, see Fig.~\ref{chiN}.   
It is obvious, that the RGM and the ED data for $\chi$ are in excellent
agreement.
This is consistent with our previous finding\cite{haertel08} that for the
1D Heisenberg ferromagnet the RGM data for
$\chi(T \to 0) $ and $\xi(T \to 0) $ are in perfect agreement with exact
Bethe ansatz results.
Moreover, the RGM calculations of $\chi$ for finite $N$ allow to
estimate the magnitude of the finite-size effects.
As can be clearly seen in Fig.~\ref{chiN}, 
the finite-size effects at $T \lesssim 0.6$
($T \lesssim 0.4$) for $J_2=0.1$ ($J_2=0.3$) become very large.
Since the correlation length $\xi$ becomes smaller with increasing $J_2$ (see
the inset in Fig.~\ref{xi_T}), the finite-size effects become less pronounced with
growing frustration.
Nevertheless, based on our data we argue that the finite systems of $N=16$ and
$N=20$, accessible by
exact full
diagonalization,
are not representative for the thermodynamic limit, see also
Refs.~\onlinecite{schmidt07} and \onlinecite{schmidt07_2}.
Note that for the 1D case, the ED data for $N=20$ are
in good agreement with RGM data for $N \to \infty$.\cite{haertel08}  Consistent
to that observation, in two
dimensions 
the system of $N=400$ sites with the same linear extension 
is already close to
the thermodynamic limit, see Fig.~\ref{chiN}.

Next we use the RGM for $N\to\infty$ to investigate the critical behavior 
of $\chi$ and $\xi$ for $T\to 0$ in more detail.
We start with a brief discussion of the low-temperature behavior of the unfrustrated
ferromagnet. Contrary to the 1D  case, where for $J_2=0$ exact Bethe results are
available,\cite{yamada} we have only approximate results for the
2D  model.
From low-temperature expansions of the susceptibility and the correlation
length  for
the model with $J_2=0$ using renormalization group approaches,\cite{kopietz,karchev}
the modified spin-wave threory,\cite{takahashi} and the RGM\cite{SSI94}
it is known that 
for $T\to 0$ the susceptibility behaves as $\chi  \propto 
T^s\exp\left(b/T\right)$ and the 
correlation length as
 $\xi \propto T^\sigma\exp\left(\beta/T\right)$.  While
different leading exponents $s$ and $\sigma$ for the (less important) 
preexponential factor
were obtained by different methods,     
the exponential divergence is obtained by all
authors.\cite{takahashi,SSI94,kopietz,karchev}
However, different values for the coefficients $b$ and $\beta$ were reported,
namely,
$b(J_2=0)=\pi/2$ using RGM,\cite{SSI94}  
  $b(J_2=0)=\pi$ using modified spin-wave
theory\cite{takahashi} or renormalization group approach,\cite{kopietz} and 
$b(J_2=0)=0.1327\pi$ using a different version of the renormalization-group
approach.\cite{karchev}
In all these papers\cite{takahashi,SSI94,kopietz,karchev} it was found that the 
coefficients in the exponents fulfill the relation $\beta=b/2$.
We mention further, that early numerical studies  based on quantum Monte-Carlo
calculations (using, however, $\chi$  and $\xi$ data only for quite large
temperatures)  
give $b(J_2=0)\approx 4.5 \approx 1.43 \pi $ (Ref.~\onlinecite{chen91})  and
$\beta(J_2=0)=0.254\pi$ (Ref.~\onlinecite{manousakis88}). 
Let us finally argue, that the results for 
the coefficient $b$ obtained by modified spin-wave
theory\cite{takahashi} as well as renormalization group
approach\cite{kopietz} and by the RGM\cite{SSI94} seem to be most reliable,
since these methods are well-tested.
Moreover, 
for the 1D spin-$1/2$ Heisenberg ferromagnet
it was shown that the RGM\cite{haertel08,SSI94} as well as the modified
spin-wave
theory\cite{takahashi} reproduce the exact Bethe-ansatz results for the
low-temperature behavior of $\chi$ and $\xi$.
Hence, it is to some extent surprising,
that there is a difference  by a factor of
2 in the coefficient $b$ between the value obtained by modified spin-wave
theory\cite{takahashi} (as well as renormalization group
approach\cite{kopietz}) and the RGM for the 2D unfrustrated ferromagnet.
This discrepancy is a known
but unresolved problem.\cite{SSI94,ihle_new}
However, we believe that our general conclusions  concerning the exponential divergence
of the frustrated model, see the discussion below, 
are not affected by this problem.

Next we use 
the RGM results
to determine the
coefficients $b$  and $\beta$
as functions of $J_2$. We assume that the general 
low-$T$ behavior of these quantities is preserved
for $0<J_2<J^c_2$, cf. Ref.~\onlinecite{haertel08}, and
fit our numerical RGM 
data for $\chi$ and $\xi$ at low
temperatures  using the ansatzes
$  
\chi =\left(a_0 T^{-1}+a_1+a_2 T\right)\exp\left(\frac{b}{T}\right)
$
and
$
  \xi =\sqrt{\left(\alpha_0 T^{-1}+\alpha_1 +\alpha_2
T\right)}\exp\left(\frac{\beta}{T}\right)
$.
Note that the leading power $T^{-1}$  in the preexponential
functions was derived for the unfrustated case with the RGM in
Ref.~\onlinecite{SSI94}.
We consider $b$, $a_0$, $a_1$, and $a_2$ as well as    
$\beta$, $\alpha_0$, $\alpha_1$, and $\alpha_2$
as independent fit parameters. 
Recall that we can calculate RGM data only for temperatures down to
a certain $T_0$, where $T_0$
(e.g., $T_0=0.161$ for $J_2=0$ and $T_0=0.052$ for
$J_2=0.4$)
is
the  lowest temperature where the system of RGM equations converges (see
Sec.~\ref{RGM}). Thus, in addition to the leading power
$T^{-1}$   in the preexponential
function 
it is reasonable to consider higher order terms
to achieve an optimal fit of the RGM data.  
For the fit we use 500 equidistant data points in the interval $T_0\ldots
T_0+T_{cut}$, where
$T_{cut}$ is set to $0.05$.
The fit of the numerical RGM data reproduces the analytical
results  of Ref.~\onlinecite{SSI94} for $J_2=0$, i.e.,
$b(J_2=0)=2\beta(J_2=0)=\pi/2$,
with  a precision of four digits. 

After having tested our fitting procedure by comparison with the analytical
predictions for $J_2=0$, 
we now consider the frustrated model, 
where (to the best of our  knowledge) no other results are available. 
From our numerical data for $\chi$ and $\xi$ we determine the $J_2$ dependence of the 
coefficients $b$ and  $\beta$.
We find that 
the numerical data for $b$ and $\beta$ obtained by
the fitting procedure described above 
are very well described by 
a linear decrease in both parameters with increasing frustration,
\begin{equation} \label{b_beta}
  b=2\beta=-\frac{\pi}{2}\left(J_1+2J_2\right) .
\end{equation}
Obviously, both parameters
would be zero at the classical transition point $J_2=0.5$,
but they are still finite at the transition point $J^c_2 \approx 0.44$ of the
quantum model. Hence, the exponential divergence  is present in the full
parameter range $J_2 \le J^c_2$ where the ground state is ferromagnetic.  
We emphasize that this result is contrary to  
the behavior observed for the 
1D frustrated spin-$1/2$ ferromagnet, where
the critical properties change at the zero-temperature transition point.\cite{haertel08} 
We mention further that  
the leading coefficients $a_0$
and $\alpha_0$ of the preexponential functions, see the expressions for
$\chi$ and $\xi$ given above,   
vanish at the transition point $J^c_2$, whereas
the next coefficients  $a_1$ and 
$\alpha_1$ remain finite at $J^c_2$.
Hence, the preexponential temperature dependence is changed approaching the
zero-temperature transition.

Let us mention, that the linear decrease  in the
coefficients $b$ and $\beta$, see Eq.~(\ref{b_beta}),
found by fitting the low-temperature behavior of $\chi$ and $\xi$
is the same as that obtained analytically for the zero-temperature spin-stiffness
$\rho_s$, see
Sec.~\ref{RGM}.  This relation between $\rho_s$ 
and the divergence
of the correlation length and the susceptibility is in accordance with general arguments\cite{ivanov,kopietz91}
concerning the low-temperature physics of low-dimensional Heisenberg
ferromagnets.

Another interesting quantity is the specific heat $C_V$ 
shown in Fig. \ref{specheat}. For $J_2=0$ the specific heat exhibits 
a typical broad maximum at about $T=0.562$. 
Increasing the frustration
the height of this maximum becomes smaller and it is shifted to lower
temperatures.  
\begin{figure}[ht]
  \begin{center}
    \includegraphics[height=6cm]{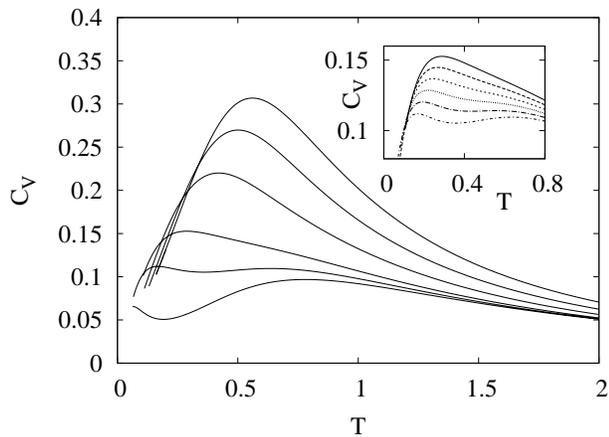}
  \end{center}
  \caption{Specific heat calculated by 
RGM for $J_2=0,0.1,0.2,0.3,0.35$, and $0.4$, from top to bottom. The inset shows the specific heat
for $J_2=0.3$ to $0.35$ in 
steps of $0.01$.}\label{specheat}
\end{figure}
Interestingly, within our RGM approach the shape of the $C_V(T)$ curve
changes for large frustration. For $J_2\gtrsim 0.34$ the specific heat shows
two maxima, one at low temperatures ($T<0.193$) and another one at high
temperatures ($T>0.6$). 
This extra low-temperature maximum signals the
emergence  of an additional  low-energy scale 
when approaching the transition point $J^c_2$. Then many low-lying multiplets appear   above the
fully polarized ferromagnetic ground-state multiplet.
Hence, 
the appearance of the additional low-temperature maximum in $C_V(T)$ can be
attributed to a subtle interplay between all of these low-lying
states.
Note that a similar behavior was found 
for the 1D case.\cite{haertel08}

\section{Summary}\label{summary}
In this paper we investigated the influence of the frustrating  
NNN coupling $J_2 < J^c_2$ on the thermodynamic 
properties of the 2D spin-1/2 Heisenberg ferromagnet using the
spin-rotation-invariant Green's function method (RGM). 
The RGM estimate for the critical frustration $ J^c_2$, where
the ferromagnetic ground state breaks down, is $J^c_2 \approx 0.44|J_1|$.

We tested the method by comparing RGM data for the spin-spin correlation
functions and the uniform susceptibility $\chi$ calculated for finite
lattices of $N=16$ and $N=20$ sites with corresponding 
exact-diagonalization data for the same lattice sizes and found a good 
agreement between both methods.
However, the comparison of RGM data for finite lattices and for $N\to \infty$ 
indicates strong finite-size effects at low temperatures.
This leads to  the conclusion  that
the finite systems of $N=16$ and $N=20$ 
are not representative for the low-temperature thermodynamics 
of large systems.

As it is known from the Mermin-Wagner theorem\cite{mermin66}
the thermal fluctuations are strong enough to suppress magnetic LRO for the
Heisenberg ferromagnet in dimension $D<3$ at any finite temperature. Due to
frustration the fluctuations are further enhanced. Thus, frustration  leads to a significant
suppression of magnetic correlations at finite temperatures even if the
ground state remains ferromagnetic.

The low-temperature behavior 
of the susceptibility $\chi$ and the correlation length 
$\xi$ in the ferromagnetic ground-state region $J_2 < J^c_2$ 
exhibits the exponential divergences 
$\chi \propto \exp\left(\frac{\pi\left(|J_1|-2J_2\right)}{2T}\right)$ 
and $\xi \propto \exp\left(\frac{\pi
\left(|J_1|-2J_2\right)}{4T}\right)$.
Although the numerator in the exponent decreases with growing $J_2$ 
the exponential divergence perpetuates in the whole parameter region $J_2
\le J^c_2$. 

The specific heat calculated within the RGM  
exhibits a double-maximum structure for $J_2\gtrsim 0.34$ 
analogous to the 1D model.

To test our theoretical predictions it would be interesting 
to reconsider experimentally  quasi-2D spin-1/2
ferromagnets, such as
K$_2$CuF$_4$, Cs$_2$CuF$_4$, Cs$_2$AgF$_4$, La$_2$BaCuO$_5$,
and
Rb$_2$CrCl$_4$\cite{feldkamp95,feldkamp98,manaka03,tennant05,Kasinathan06} 
with respect to a possible frustration and its influence on low-temperature
thermodynamics.\\

{{\it \bf  Acknowledgments:}}
This work was supported by the DFG (project RI615/16-1 and DR269/3-1).
For the exact diagonalization  J.~Schulenburg's {\it
spinpack} was used. J.R. thanks N.B. Ivanov for useful discussions.
}


\begin{thebibliography}{99}
\bibitem{diep04}
{\it Frustrated Spin Systems},
H.~T.~Diep, Ed.
(World Scientific, Singapore, 2004).

\bibitem{002}
{\it Quantum Magnetism},
U.~Schollw\"{o}ck, J.~Richter, D.~J.~J.~Farnell, R.~F.~Bishop, Eds.
(Lecture Notes in Physics, {\bf 645})
(Springer, Berlin, 2004).
\bibitem{j1j2} P.~Chandra and B.~Doucot,  { Phys. Rev. B} {\bf 38},
  9335 (1988); H.J.~Schulz and T.A.L.~Ziman, { Europhys. Lett.}
  {\bf 18}, 355 (1992);
H.J. Schulz, T.A.L.~Ziman, and D.~Poilblanc { J.~Phys.~I}
  {\bf 6}, 675 (1996);
J.~Richter, N.B.~Ivanov, and K.~Retzlaff,
      Europhys. Lett. {\bf 25}, 545 (1994);
A.~E.~Trumper, L.~O.~Manuel, C.~J.~Gazza, and H.~A.~Ceccatto,
Phys. Rev. Lett. {\bf 78}, 2216 (1997); 
L.~Capriotti, F.~Becca, A.~Parola, and S.~Sorella, 
{ Phys. Rev. Lett.} {\bf 87}, 097201 (2001);
R.R.P.~Singh, Weihong Zheng , J.~Oitmaa, 
O.P.~Sushkov, and C.J.~Hamer,
{ Phys. Rev. Lett.} {\bf 91}, 017201 (2003);
J. Sirker, Z. Weihong, O. P. Sushkov, and J. Oitmaa,
Phys. Rev. B {\bf 73}, 184420 (2006);
R.~Darradi, O.~Derzhko, R.~Zinke, J.~Schulenburg,  S.~E.~Kr\"uger, and
       J.~Richter,
       Phys. Rev. B {\bf 78}, 214415 (2008);
V. Murg, F. Verstraete, and J. I. Cirac, Phys. Rev. B {\bf 79},
195119 (2009), J. Richter and J. Schulenburg, Eur.
Phys. J. B {\bf 73}, 117 (2010).

\bibitem{melzi} 
R.~Melzi, P.~Carretta, A.~Lascialfari, M.~Mambrini, M.~Troyer, 
P.~Millet, and F.~Mila, Phys. Rev. Lett. \textbf{85}, 1318 (2000); R.~Melzi, S.~Aldrovandi, 
F.~Tedoldi, P.~Carretta, P.~Millet, and F.~Mila, Phys. Rev. B \textbf{64}, 024409
(2001);
H.~Rosner, R.R.P.~Singh, W.H.~Zheng, J.~Oitmaa, and W.E.~Pickett, Phys. Rev. B \textbf{67} 014416
(2003).
\bibitem{kaul04} E.~E.~Kaul, H.~Rosner, N.~Shannon, R.V.~Shpanchenko, and C.~ Geibel, 
J. Magn. Magn. Mater. \textbf{272-276(II)}, 922 (2004).
\bibitem{jmmm07} M. Skoulatos, J.P. Goff, N. Shannon, E.E. Kaul, C. Geibel, A.P. Murani, M.
Enderle, and  A.R. Wildes
J. Magn. Magn. Mater. \textbf{310}, 1257 (2007).
\bibitem{carretta2009}
P. Carretta,  M. Filibian,  R. Nath,  C. Geibel, and P. J. C. King,
Phys. Rev. B {\bf 79}, 224432 (2009).
\bibitem{enderle}
M. Skoulatos, J.P. Goff, C. Geibel, E.E. Kaul, R. Nath, N. Shannon,
B. Schmidt, A.P. Murani, P.P. Deen, M. Enderle, and A.R. Wildes,
Europhys. Lett. {\bf 88}, 57005 (2009).

\bibitem{kageyama05} H.~Kageyama, T.~Kitano, N.~Oba, M.~Nishi, S.~Nagai, K.~Hirota, L.~Viciu, J.B.~Wiley, J.~Yasuda, Y.~Baba, Y.~Ajiro, 
and K.~Yoshimura, J. Phys. Soc. Jpn. \textbf{74}, 1702 (2005).

\bibitem{rosner09}
A.A. Tsirlin and H. Rosner,
Phys. Rev. B {\bf 79}  214417  (2009).
\bibitem{rosner09a}
 A.A. Tsirlin, B. Schmidt,  Y. Skourski,
   R. Nath, C. Geibel, and H. Rosner,
Phys. Rev. B {\bf 80} 132407  (2009).
\bibitem{nath2008}
R. Nath, A.A. Tsirlin, H. Rosner, and C. Geibel,
Phys. Rev. B {\bf 78}  064422 (2008).
\bibitem{shannon04} N.~Shannon, B.~Schmidt, K.~Penc, and P.~Thalmeier, Eur. Phys. J. B \textbf{38}, 599
(2004).
\bibitem{shannon06} N.~Shannon, T.~Momoi, and P.~Sindzingre, Phys. Rev. Lett. \textbf{96}, 027213
(2006).
\bibitem{sindz07}
P. Sindzingre, N. Shannon and T. Momoi, J. Magn. Magn. Mat.
{\bf  310}, 1340 (2007).
\bibitem{schmidt07} B.~Schmidt, N.~Shannon, and P.~Thalmeier, J. Phys. Cond. Mat. \textbf{19}, 145211
(2007).
\bibitem{schmidt07_2} B.~Schmidt, N.~Shannon, and P.~Thalmeier, J. Magn. Magn. Mater. \textbf{310}, 1231
(2007).
\bibitem{sousa}  J.R.~Viana and J.R.~de~Sousa,
 Phys.\ Rev.\ B {\bf 75}, 052403 (2007). 
\bibitem{shannon09} P. Sindzingre, L. Seabra, N. Shannon, and
T. Momoi,
J. Phys.: Conf.  Series {\bf 145}, 012048 (2009).
 \bibitem{momoi}
 R. Shindou and T. Momoi,
  Phys.\ Rev.\ B {\bf 80}, 064410 (2009).
\bibitem{schulen} J. Richter, R. Darradi, J.~Schulenburg,
D.J.J. Farnell, and H. Rosner, arXiv:1002.2299 (2010).
\bibitem{dmitriev96}
D.~V.~Dmitriev, V.~Ya.~Krivnov, and A.~A.~Ovchinnikov, Phys. Rev. B
\textbf{55}, 3620 (1997). 
 


\bibitem{feldkamp95} S.~Feldkemper, W.~Weber, J.~Schulenburg, and J.~Richter, Phys. Rev. B \textbf{52}, 313
(1995).
\bibitem{feldkamp98} S.~Feldkemper and W.~Weber, Phys. Rev. B \textbf{57}, 7755
(1998).
\bibitem{manaka03} H.~Manaka, T.~Koide, T.~Shidara, and I.~Yamada, 
Phys. Rev. B \textbf{68}, 184412 (2003).
\bibitem{tennant05}
S.E.McLain, D.A.Tennant, J.F.C.Turner,
    T. Barnes, M.R. Dolgos, Th. Proffen , B.C. Sales,
    and R.I Bewley, 
arXiv:cond-mat/0509194.
\bibitem{Kasinathan06} 
D. Kasinathan, A.B. Kyker,  and D.J. Singh
Phys. Rev. B {\bf 73}, 214420 (2006).

\bibitem{haertel08} M.~H\"{a}rtel, J.~Richter, D.~Ihle, 
and S.-L.~Drechsler, Phys. Rev. B \textbf{78}, 174412 (2008);
J. Richter, M. H\"{a}rtel, D. Ihle, and  S.-L. Drechsler,
       J.Phys.:Conf.Series {\bf 145}, 012064 (2009).
\bibitem{takahashi}
M.~Takahashi, Prog. Theo. Phys. Supp. \textbf{87}, 233 (1986); M.~Takahashi, Phys. Rev. Lett. {\bf 58} 168
(1987). 
 \bibitem{kopietz}
 P.~Kopietz and S. Chakravarty,
 Phys. Rev. B {\bf 40}  4858
 (1989).
\bibitem{karchev}
N.~Karchev,
Phys. Rev. B {\bf 55}  6372
(1997). 
\bibitem{manousakis88} E.~Manousakis and R.~Salvador, Phys. Rev. B \textbf{39}, 575
(1989).
\bibitem{chen91} Y.C.~Chen, H.H.~Chen, and F.~Lee, Phys. Rev. B \textbf{43}, 11082
(1991).
\bibitem{timm2000} P. Henelius,  A.W. Sandvik,
C. Timm, and  S. M. Girvin
Phys. Rev. B {\bf 61}, 364 (2000).
\bibitem{ihle_new} I. Juh\'{a}sz Junger, D. Ihle, L. Bogacz, and W. Janke, Phys. Rev. B {\bf 77}, 174411 (2008). 
\bibitem{kondo} J. Kondo and K. Yamaji, Prog. Theor. Phys. \textbf{47}, 807
(1972); H. Shimahara and S. Takada, J. Phys. Soc. Jpn. \textbf{60}, 2394
(1991); S. Winterfeldt and D. Ihle, Phys. Rev. B \textbf{56}, 5535 (1997).
\bibitem{SSI94} F.~Suzuki, N.~Shibata, and C.~Ishii, J. Phys. Soc. Jpn. \textbf{63}, 1539 (1994).
\bibitem{magfeld} I. Junger, D. Ihle, J. Richter, and A. Kl\"{u}mper, Phys. Rev. B \textbf{70}, 
104419 (2004).
\bibitem{antsyg} T.N.~Antsygina,
M.I.~Poltavskaya, I.I.~Poltavsky, and K.A.~Chishko, Phys. Rev. B {\bf 77}, 024407 (2008).
\bibitem{elk} W. Gasser, E. Heiner and K.~Elk,  {\it Greensche Funktionen in Festk\"{o}rper- und Vielteilchenphysik} 
(Wiley, Berlin 2001).
\bibitem{izumov02}
Yu. A. Izyumov, N.I. Chaschin, and V. Yu. Yushankhai, Phys. Rev. B  {\bf 65}, 214425 (2002).
\bibitem{froebrich06}P. Froebrich  and P.J. Kuntz,
Physics Reports {\bf 432}, 223 (2006). 
\bibitem{rgm_new} W. Yu and S. Feng, Eur. Phys. J. B {\bf{13}}, 265 (2000); 
B.H. Bernhard, B. Canals, and C. Lacroix, Phys. Rev. B {\bf{66}},
104424 (2002); D. Schmalfu{\ss}, J. Richter, and D. Ihle, Phys. Rev. B \textbf{70}, 184412 
(2004);
D. Schmalfu{\ss}, J. Richter, and D. Ihle, Phys. Rev. B {\bf 72}, 224405
(2005).
\bibitem{j1j2_a}
L. Siurakshina, D. Ihle, and R. Hayn, Phys. Rev. B \textbf{64}, 104406
(2001);
D. Schmalfu{\ss}, R. Darradi, J. Richter, J.~Schulenburg,  and D.~Ihle, 
       Phys. Rev. Lett. {\bf 97}, 157201 (2006). 
\bibitem{ihlefinite} I. Juh\'{a}sz Junger, D. Ihle, and J. Richter, Phys. Rev. B {\bf 72}, 064454
(2005).
\bibitem{NR3} W.H.~Press, S.A.~Teukolsky, W.T.~Vetterling, B.P.~Flannery, 
{\it Numerical Recipes in C++, The Art of Scientific Computing} (Cambridge University Press, Cambridge,
2007).
\bibitem{yamada} M. Yamada  and M. Takahashi, 
J. Phys. Soc. Jpn. \textbf{55}, 2024 (1986); M. Yamada, 
J. Phys. Soc. Jpn. \textbf{59}, 848 (1990).

\bibitem{kopietz91}
 P.~Kopietz and G. Castilla, Phys. Rev. B {\bf 43}  11100 (1991).

\bibitem{ivanov}   N.B. Ivanov, Condens. Matter Phys. (L'viv) {\bf 12}, 435 (2009) (see also arXiv:0909.2182).


\bibitem{mermin66}
     N. Mermin and H. Wagner, Phys. Rev. Lett. {\bf 17},
     1133 (1966).


\end{thebibliography}
\end{document}